\documentstyle[emulateapj,danonecolfloat]{article}

\newcommand{\D}{$D$}
\newcommand{\Done}{$D$1}
\newcommand{\Dtwo}{$D$2}

\newcommand{\cobedmr}{{{\sl COBE\/}/DMR}}
\newcommand{\sk}{{\sl SK\/}}
\newcommand{\qmap}{{\sl QMAP\/}}
\newcommand{\mat}{{\sl MAT\/}}
\newcommand{\tocoA}{{\sl TOCO97\/}}
\newcommand{\tocoB}{{\sl TOCO98\/}}

\slugcomment{Submitted 1999 June 23 to Astrophysical Journal Letters}

\lefthead{Miller et al.}
\righthead{TOCO98}

\begin{document}

\twocolumn[

\title{A Measurement of the Angular Power Spectrum of the CMB from $\ell =
100~{\rm to}~400$}

\author{A. D. Miller\altaffilmark{1},
R. Caldwell\altaffilmark{1,2},
M. J. Devlin\altaffilmark{2},
W. B. Dorwart\altaffilmark{1},
T. Herbig\altaffilmark{1},
M. R. Nolta\altaffilmark{1}, 
L. A. Page\altaffilmark{1},
J. Puchalla\altaffilmark{2}, 
E. Torbet\altaffilmark{1},
H. T. Tran\altaffilmark{1}}

\begin{abstract}

We report on a measurement of the angular spectrum of the CMB between
$l\approx 100$ and $l\approx 400$ made at 144~GHz from
Cerro Toco in the Chilean altiplano.
When the new data are combined with previous data at 30 and 40~GHz,
taken with the same instrument observing the same section of sky,
we find: 1) a rise in the angular spectrum to a maximum with
$\delta T_l \approx 85~\mu$K at $l\approx 200$ and a fall at $l>300$,
thereby localizing the peak near $l\approx 200$; and 2) that the anisotropy
at $l\approx 200$ has the spectrum of the CMB.

\end{abstract}


]
\altaffiltext{1}{Princeton University, Physics Department, Jadwin Hall, Princeton, NJ 08544}\altaffiltext{2}{University of Pennsylvania, Department of Physics and Astronomy, David Rittenhouse Laboratory, Philadelphia, PA 19104}

\section{Introduction}

It is widely recognized that the characterization of the cosmic microwave
background (CMB) anisotropy
is essential for understanding the process of cosmic structure formation 
(e.g. \cite{whu97}, \cite{ben97}, \cite{tt99}).
If some of the currently popular models prove correct, the anisotropy
may be used to strongly constrain cosmological parameters
(e.g. \cite{jung95}, \cite{bond98}).
Summaries of the state of our knowledge of the CMB (e.g. \cite{bjk98}
(BJK), \cite{pw99}) suggest the existence of a peak in the angular
spectrum near $l=200$.
In particular, BJK show $150 \le l_{\rm peak}\le 350$. 
Since their analysis there have been additional results at $l>200$ 
that lend support to their picture
(\cite{bak99} ({\sl CAT\/}), \cite{glanz99} ({\sl VIPER\/}),
\cite{wilson99} ({\sl MSAM\/})).
Here, we report the results from the \tocoB\ campaign of the Mobile
Anisotropy Telescope (\mat) which probes from $l\approx 100$ to
$l\approx 400$.

\section{Instrument}
\label{inst}

The \mat\ telescope, based on the design in \cite{wol97}, 
is described briefly in \cite{torb99} and \cite{dev98} and is documented
on the web\footnote{Details of the experiment, synthesis vectors,
likelihoods, data, and analysis code may be found at
http://www.hep.upenn.edu/CBR/ and
http://physics.princeton.edu/{\char'176}cmb/.}.
In this paper, we focus on results
from the two \D-band (144~GHz) channels. The receivers use
SIS mixers designed and fabricated by A.~R.~Kerr and S.-K.~Pan 
of NRAO  and A.~W.~ Lichtenberger of
the University of Virginia (\cite{ker93}). The six other detectors 
in the focal plane are 30 and 40
GHz high electron mobility transistor (HEMT) amplifiers designed by M. Pospieszalski
(\cite{posp92}, \cite{posp94}).

%(National Radio Astronomy Observatory)

The mixers, which operate in double sideband mode, are fed with a 
25\%~bandwidth corrugated feed cooled to 4.5~K.
The 144~GHz local oscillator (LO) is cavity stabilized and thermally controlled.
The cryogenic IF HEMT amplifier, which operates between 4 and 6 GHz, is also of NRAO design.
The resultant passband has been measured (\cite{rob96}) to be approximately 138-140
and 148-150~GHz.
The total system sensitivity (including atmospheric loading) for each
receiver is $\approx 1.3~{\rm mK\,s}^{1/2}$ (Rayleigh-Jeans) with the
SIS body operating at $\approx 4.4$~K.

The \Done\ feed ($az=207\fdg47$, $el=40\fdg63$ at the center of the chop)
is near the center of the
focal plane, resulting in $\theta_{\rm FWHM}\approx0\fdg2$
($\Omega_{\rm D1}=1.36\times 10^{-5}$~sr) while \Dtwo\ is displaced
from the center by 2.9~cm ($az=205\fdg73$, $el=40\fdg13$),
resulting in $\theta_{\rm FWHM}\approx0\fdg3$
($\Omega_{\rm D2}=2.93\times 10^{-5}$~sr).
\Done\ is polarized with the $E$-field in the horizontal direction and
\Dtwo\ with the field in the vertical direction.
No use is made of the polarization information in this analysis.

In the 1997 campaign (\cite{torb99}), a microphonic coupling rendered
the \D-band data suspect.
The problem was traced to a combination of the azimuth drive motor and
the chopper. The coupling was effectively eliminated for the 1998 campaign.
In addition, the chopper amplitude was reduced from 2\fdg96 to 2\fdg02
and the frequency reduced from 4.6~Hz to 3.7~Hz.
In all other respects, the instrument was the same as for 1997. 

The telescope pointing is established through observations of
Jupiter and is monitored with two redundant encoders on
both the azimuth bearing and on the chopper. The absolute errors
in azimuth and elevation are 0\fdg04, and the relative errors are 
$<0$\fdg01. The chopper
position, which is calibrated in the field, is sampled 80 times per
chop.  When its {\it rms} position over one cycle deviates more than
0\fdg015 from the average position (due to wind loading), we reject the data.

The analysis uses data between 20 and 200 Hz. These frequencies are well
removed from the refrigerator cycle frequency at 1.2 Hz, the chopper frequency,
and the Nyquist frequency at 592 Hz. The amplitude of the electronic transfer
function varies by $<2$\% over this band.

\section{Observations and Calibration}
\label{obs}

Data were taken at a 5200~m site\footnote{ The Cerro Toco site of
the Universidad Cat\'olica de Chile was made available through the
generosity of Prof. Hern\'an Quintana, Dept. of Astronomy and
Astrophysics. It is near the proposed MMA site.} on the side of
Cerro Toco ({\it lat\/} = -22\fdg95, {\it long\/} = 67\fdg775),
near San Pedro de Atacama, Chile, from Aug. 26, 1998 to Dec. 7, 1998.
For the anisotropy data, the primary optical axis is fixed at
{\it az} = 207\fdg41, {\it el} = 40\fdg76, $\delta$ = -60\fdg9 and the
chopper scans 6\fdg12 of sky.
We present here the analysis of data from Sept. 3, 1998 to Oct. 28, 1998.

%
% \centerline{\vbox{\epsfxsize=3.75in\epsfbox{toco98_coverage.ps}}}
% \noindent{\small Fig 1. 
% The region of sky over which we observe superimposed
% on the IRAS dust map at 100 microns. A scan line is plotted for each
% channel for 15 min intervals. Though the whole ring is observed 
% ($\approx ??$ sq deg),
% only the section marked with a line is used for anisotropy analysis
% ($\approx ??$ sq. deg). The SCP is in the center, the LMC is
% at ra $= 82^{\circ}$ and dec $= -70^{\circ}$ \label{fig-1}}
% \smallskip
%
%
% \begin{figure*}
% \plotone{toco98_coverage.ps}
% \caption{The region of sky over which we observe superimposed
% on the IRAS dust map at 100 microns. Though the whole ring is observed,
% only the section marked with a line is used for anisotropy analysis. \label{fig-1}}
% \end{figure*}
%

Jupiter is used to calibrate all channels and map the beams. Its
brightness temperature is 170 K in \D-band
(\cite{gri86}, \cite{uli81}) and the intrinsic
calibration error is $ 5\%$. We account for the variation
in angular diameter. To convert to thermodynamic units
relative to the CMB, we multiply by
$1.67\pm0.03$. The error is due to incomplete knowledge of the passbands.

After determining the beam parameters from a global
fit of the clear weather Jupiter calibrations, the standard deviation in
the measured solid angle is 5.5\% for \Done\ and 4\% for \Dtwo. 
Jupiter is observed on average within 2 hours of the prime observing
time (approximately 10 PM to 10 AM local). The responsivity varies $\approx20$\% 
over two months. In all, there are 
$\approx 35$ Jupiter calibrations in each channel.

%
% After determining the beam parameters from a global
% fit of the clear weather Jupiter calibrations, the standard deviation in
% the measured solid angle is 5.5\% for \Done\ and 4\% for \Dtwo. 
% Typically, the calibration amplitudes are constant during the roughly eight days
% between maintenance shut-downs. After the refrigerator power has been
% cycled, the calibration may 
% change by up to 20\%. There is also a general long term drift. 
% Jupiter calibrations occur within three hours of the prime
% observing hours (10 PM to 10 AM local). In all, there are 
% $\approx 35$ calibrations in each channel.
%
%
%
% Deemed not significant enough to put in:
% Occasionally, there is some contamination in D1 due to
% water vapor accumulation at the center of the vacuum window. This is
% evident as a slight alteration in the beam profile though not in the
% calibrations amplitude (is this correct??)
%
%
% MD has soem different words for the middle. ``The Jupiter and pulse
% calibrations agree to XXX ove XXX.''
%
%
% The pulse height is correlated to the temperature of the coldhead
% allowing us to bound variations in the gain.
%
%

To verify the calibration between observations of Jupiter,
a 149 GHz tone is coupled to the detectors through the LO port
for 40~msec every 200~seconds. Its effective temperature is $\approx1~$K.  
There is good long term agreement between the Jupiter and pulse calibrations.
The short-term ($<$ 1 day) calibration is determined with a fit of the pulses
to the Jupiter calibrations. The measurement uncertainty in the calibration
is~7\%. 

%
%In addition, we ran the full analysis pipeline using only the Jupiter 
%calibrations with interpolating and found no difference in the results. 
%

The total 1$\sigma$ calibration error of 10\% for \Done\ and 9\% for
\Dtwo\ is obtained from the quadrature sum of the above sources.
In the full analysis, \Done\ and \Dtwo\ are combined; thus the uncorrelated 
component of the error adds in quadrature yielding an error for the
combination of 8\%. 

\section{Data Reduction}

The data reduction is similar to that of the \tocoA\ experiment
(\cite{torb99}).
We use the terminology discussed there and in \cite{net97},
though we now use Knox filters (\cite{knox99}) (in place
of window functions) to determine the $l$-space coverage.
For \Done\ we form the 2-pt through 16-pt synthesized beams
and for \Dtwo, the 2-pt through 17-pt synthesized beams.
In practice, atmospheric contamination precludes
using the 2-pt through 4-pt data and the achieved
sensitivity renders the 17-pt and higher uninteresting.
If there is atmospheric contamination in the 5-pt data,
its level is under the 1 sigma error.
For the 7-pt and higher, atmospheric contamination is negligible.

The phase of the time ordered data relative to the beam position is determined
with observations of Jupiter and the Galaxy. In the analysis, we use the
phase for each harmonic obtained when the quadrature signal from the Galaxy is
minimized.

A quantity useful in assessing sensitivity to the beam shape
is $l_{\rm eff}\,\theta_{\rm FWHM}$. For \sk\ at $l_{\rm eff}=256$,
$l_{\rm eff}\,\theta_{\rm FWHM}= 2.5$.
For \tocoB\ at $l_{\rm eff}=409$, $l_{\rm eff}\,\theta_{\rm FWHM} = 1.5$
for \Done\ and 2.2 for \Dtwo. This corresponds to a separation
of lobes in the synthesized beam of $2\,\theta_{\rm FWHM}$ for
\Done\ and $1.3\,\theta_{\rm FWHM}$ for \Dtwo.

%
% TOCO97:
% SK: 3.04
% Python: 3.42
% D1 separation is 0.385 deg for 16 pt
% D2 separation is 0.365 deg for 17 pt
%

As with \tocoA, the harmonics are binned according to the right ascension
at the center of the chopper sweep.
The number of bins
depends on the band and harmonic as shown in Table 1. For each 
night, we compute the variance and mean of the 
data corresponding to a bin. These numbers are
averaged over the 25 good nights and used in the
likelihood analysis. 

After cuts based on pointing, the data are selected according to the
weather. For each n-pt data set, we find the mean {\it rms} of
6.5 sec averages over 15 minute segments. When this value exceeds
1.2 of the minimum value for a given day, the data from that 15 min
segment, along with the previous and subsequent 15 min segments are cut.
The effect is to keep 5-10 hour blocks of continuous good data
in any day, and to eliminate transitions into periods of poor
atmospheric stability. Increasing the cut level adds 
data to the beginning and end of the prime observing time.

%
%Plots of all the data with the cuts are shown on the web.
%
%
%As a final cut, nights with less than 4.7 hours ??? of data are
%excluded. Plots of all the data with the cuts are shown on the web.
%

The stability of the instrument is assessed through internal
consistency checks and we examine it with the distribution
of the offset of each harmonic. The offset is the average of a
night of data after the cuts have been applied (the duration ranges
from 5-10~hours) and is typically of magnitude $\approx 150~\mu$K with
standard deviation $\approx 75~\mu$K. The offsets for these data were stable
over the campaign. The 
resulting $\chi^2/\nu$ is typically 1-4. For the offsets of the quadrature
signal, $\chi^2/\nu$ is typically 1-2.
The stability of the offset led to a relatively straightforward 
data reduction.

To eliminate the potential effect of slow variations in
offset, we remove the slope and mean for each night for each harmonic. This is
accounted for in the quoted result following \cite{bond91}.

\section{Analysis and Discussion}

In the analysis, we include all known
correlations inherent in the observing strategy.
In computing the ``theory covariance matrix'' (BJK) which
encodes the observing strategy, we use the measured two dimensional beam profiles.
From the data, we determine the
correlations between harmonics due to the atmosphere.
Because the S/N is only 2-5 per synthesized beam,
and the noise is correlated between beams, we work with groups
of harmonics. This is similar to band averaging, though we use the
full covariance matrix so as to include all correlations.

Table 1 gives the results of separate analyses of \Done\ and \Dtwo.
Both channels show a fall in the angular spectrum above $l=300$. 
The fact that the results agree is an important check as the receivers (other than the optics)
are independent. It is not possible to compute
\Done$-$\Dtwo\ directly from the data because of the different beam sizes. 
The eventual production of a map will facilitate the comparison.

\begin{figure}[tb]
\centerline{\epsfxsize=2.8in\epsffile{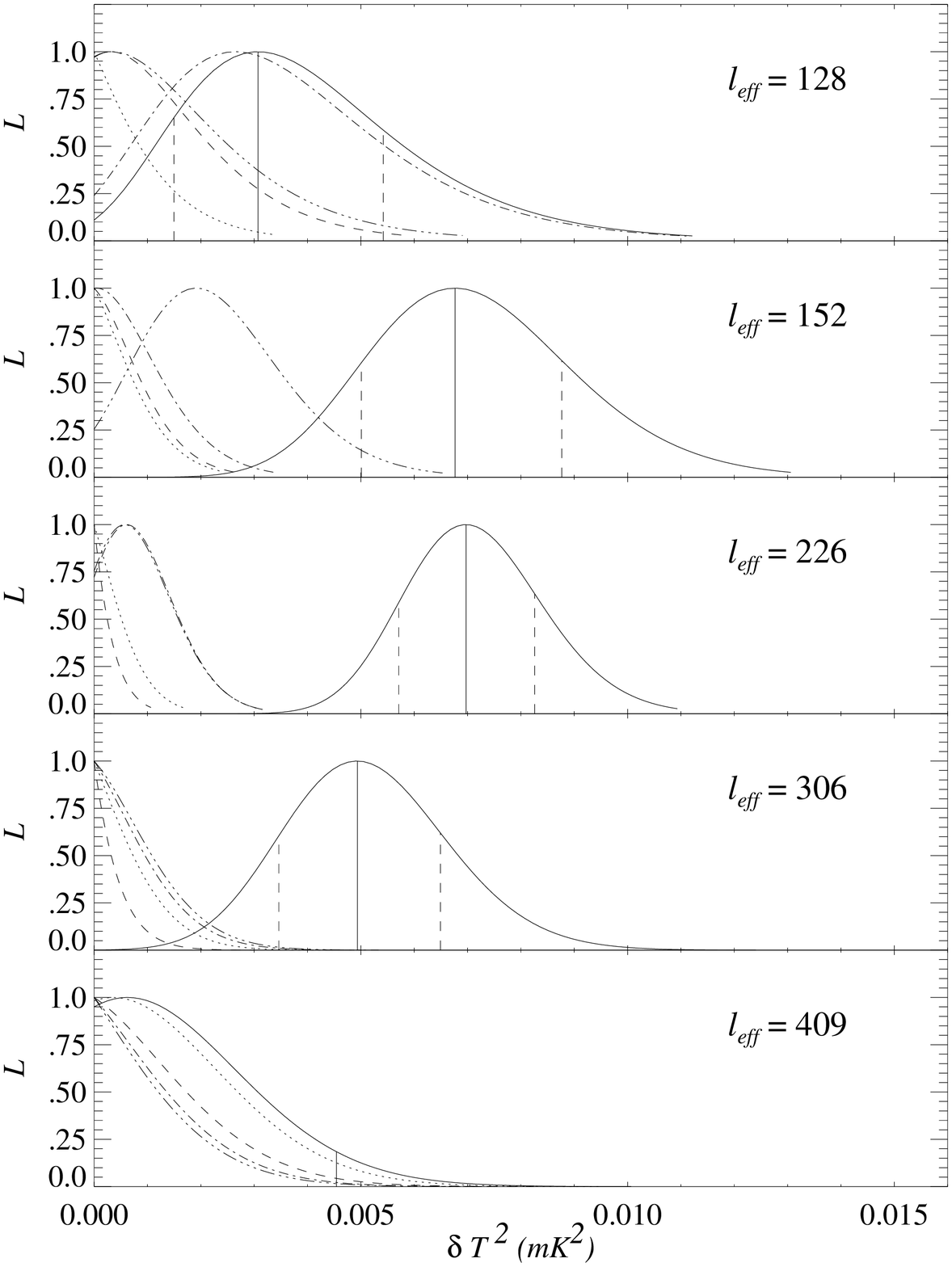}}
\vskip 2em
\caption{\footnotesize \label{fig:lh}The likelihood of the combined \Done\ and \Dtwo\ analysis (solid line) as a function of $\delta T_l^2$. The null tests: quadrature (signal with chopper sweeping one direction minus that with the chopper sweeping the other direction, dotted line), fast and slow dither (differences of subsequent $0.5\rm\,s$ and $10\rm\,s$ averages, dash and dash-dot lines respectively) and first half minus second half (dot-dot-dot-dash line), are also shown. The vertical lines indicate the maximum, $\pm 1\sigma$, or $95$\% confidence upper bound.}
\end{figure}

In the full analysis, \Done\ and \Dtwo\ are combined.
The resulting likelihoods are shown in Figure~\ref{fig:lh} along with the
results of the null tests.
Because \Done\ and \Dtwo\ observe
the same section of sky at different times, some care must be taken in
computing the correlation matrices.
The correlation coefficients between \Done\ and \Dtwo\ due to the
atmosphere are of order
$0.05$. The largest off-diagonal terms of the theory covariance matrix
are $\lesssim 0.4$.
The quoted results are insensitive to the precise values of the 
off-diagonal terms of the covariance matrix.
The combined analysis affirms what is seen in
\Done\ and \Dtwo\ individually and shows a peak in the angular 
spectrum near $l=200$.

The angular spectrum of the \tocoA\ and \tocoB\ data agree in
the regions of common $l$.
We compute the spectral index of the fluctuations by comparing band
powers. We find $\beta_{\rm CMB} = {\rm ln}(\delta T_{144}/\delta
T_{36.5})/{\rm ln}(144/36.5) = -0.04\pm 0.25$, (including
calibration error), where $\delta T_{144}$ is the weighted mean of the
two highest points for \tocoB\ and $\delta T_{36.5}$ is a similar
quantity for \tocoA\ (36.5 GHz is the average \tocoA\ frequency).
For the CMB, $\beta_{\rm CMB} = 0$. For dust,
$\beta_{\rm RJ} = 1.7$ corresponds to $\beta_{\rm CMB} = 2.05$; for free-free
emission $\beta_{\rm RJ} = -2.1$ corresponds to $\beta_{\rm CMB} = -1.75$.
Though it is possible for spinning dust grains (\cite{dl99}) to 
mimic this spectrum for our frequencies, the amplitude of this component is small
(\cite{doc98b}). In addition, the spatial spectrum of diffuse sources
like interstellar dust falls as
$l^{-3/2}$ (\cite{gau92}), so the observed peak is inconsistent with
our observations at $l\approx 100$.

The frequency spectral index $\beta_{\rm RJ}$ of unresolved extra-Galactic
sources is typically between 2 and~3, inconsistent with the measured index.
In addition, the spatial spectrum of sources rises as $\delta T_l\propto l$,
inconsistent with our observations at $l\approx 400$. Moreover, recent
analyses (e.g. \cite{teg99}) estimate the level of point source
contamination to be much lower than the fluctuations we observe. We therefore
conclude that the source of the fluctuations is the CMB.

%
% Original
%
%We assess the statistical significance of the decrease for $\delta
%T_l>300$ by comparing the likelihood distributions at $l=248$, for which
%$\delta T_l=83~\mu$K, and $l=415$. These
%two $\delta T_l$ are effectively uncorrelated.
%The probability (integrated likelihood) that $\delta T_{415} \ge 83~\mu$K is 
%0.004. In addition, the data point at $l=317$ is $1\sigma$ below the 
%value at $l=248$. In other words, we observe a decrease in power from the peak 
%value at the $>99$\% confidence level.
%
%
%We assess the statistical significance of the decrease in $\delta T_l$
%for $l>300$ by comparing just the likelihood distributions at $l=248$ ($L_{248}$), for which
%$\delta T_l=83~\mu$K, and $l=415$, for which $\delta T_l<67~\mu$K (95\%). These
%two distributions are effectively uncorrelated. The integral
%of the area normalized likelihood of $L_{415}$ for values of
%$\delta T^2$ below which $L_{415}=L_{248}$ is 0.95; this is
%the probability that $\delta T_{415}<\delta T_{248}$. The point at which
%$L_{415}=L_{248}$ is also coincidentally the $2\sigma$ lower limit on
%$\delta T_{248}$ and the 95\% upper limit on $\delta T_{415}$. The
%probability that $\delta T_{415} \le 83~\mu$K (the peak of $L_{248}$) is 
%0.996. When all the data in Figure 2 are considered, these probabilities will
%increase.
%

We assess the statistical significance of the decrease in $\delta
T_l$ for $l>300$ by comparing just the likelihood distributions
at $l=226$ ($L_{226}$, Fig.~\ref{fig:lh}), for which $\delta T_l=83~\mu$K, and
$l=409$, for which $\delta T_l<67~\mu$K (95\%). These two
distributions are  effectively uncorrelated. The point at which
$L_{409} = L_{226}$ is, coincidentally, the $2\sigma$ lower limit
on $\delta T_{226}$ and the 95\% upper limit on $\delta T_{409}$.
Thus, there is a 0.97 probability that $\delta T_{226}$ is
greater than the 95\% upper limit on $\delta T_{409}$. In
addition, the probability that $\delta T_{409} \le 83~\mu$K (the
peak of $L_{226}$) is 0.996. When all the data in Figure 2 are
considered, the significance of a decrease in $\delta
T_l$ for $l>300$ will be even higher.

\begin{figure}[tb]
\centerline{\epsfxsize=3.2in\epsffile{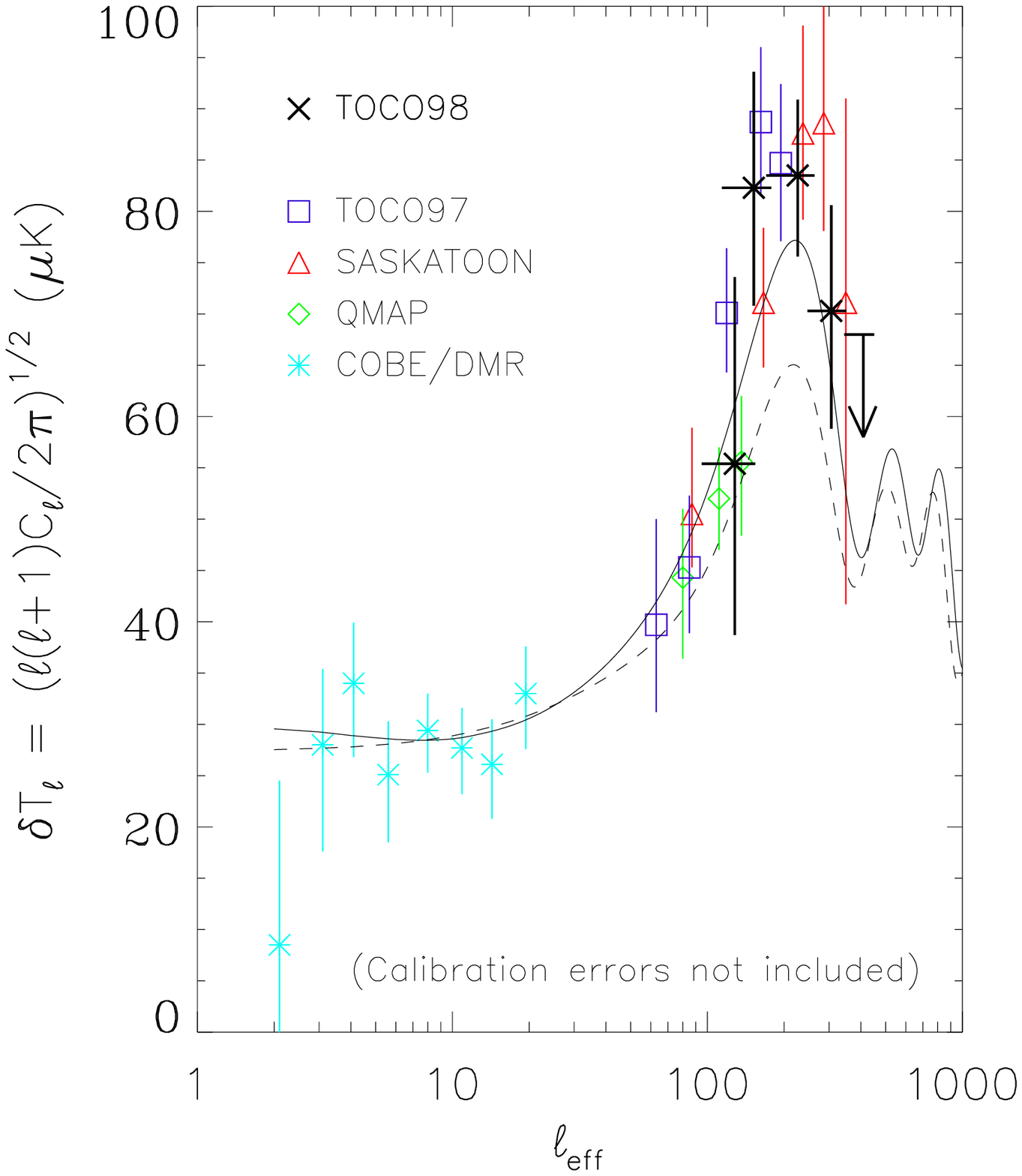}}
\caption{\footnotesize \label{fig:summary}Angular spectrum from \cobedmr, \sk, \qmap, \tocoA, and \tocoB\ \D-band. The \sk\ data have been recalibrated according to \cite{mas99}, leading to an increase of 5\%, and reduced according to the foreground contribution in \cite{doc97}, leading to a reduction of 2\% (i.e. a net 3\% increase in the mean and 5\% increase in the error bars over \cite{net97}). The revised \sk\ calibration error is 11\%. The \qmap\ data are the same as those reported in \cite{doc98} and have an average calibration error of 12\%. The correction for foreground emission is $\approx$ 2\%, though it has not yet been precisely determined and so is not included. Both \sk\ and \qmap\ are calibrated with respect to Cas-A. The \tocoA\ data, which have a calibration error of 10\%, are calibrated with respect to Jupiter. The \tocoB\ data are shown with $l$-space bandwidth as the horizontal bars. The cosmological models are computed with CMBFAST (\cite{selzal}). The dashed line is ``standard CDM'' ($\Omega_m = 1$, $\Omega_{b} = 0.05$, $h=0.5$) the solid line is a ``concordance model'' (\cite{wang99}, \cite{turn99}) with $\Omega_m=0.33$, $\Omega_{b}=0.041$, $\Omega_\Lambda=0.67$, and $h=0.65$. For \cobedmr\ we use \cite{max97}. The error bars are ``$1\sigma$ statistical.''}
\end{figure}

The weighted mean of data from \tocoA, \tocoB, and \sk\ between $l=150$ and
250, is $\overline{\delta T}_{\!\rm peak} = 82~\pm3.3\pm5.5~\mu$K (the second
error is calibration uncertainty).
This is consistent with, though slightly higher
than, the value from the \cite{wang99} concordance model plotted in
Figure~\ref{fig:summary}, which gives $\overline{\delta T}_{\!\rm peak}\approx
75\mu$K. In the context of this
model, the high $\overline{\delta T}_{\!\rm peak}$ favors a smaller
$\Omega_m h^2$ (e.g. larger ``cosmological constant'') or more baryons.

Figure~\ref{fig:summary}  shows results taken over six years and seven observing
campaigns and three different experiments. Though a detailed confrontation with cosmological
models will have to await a thorough analysis and comparison with other
experiments, a straightforward read
of the data indicates a rise to
$\delta T_{\rm peak}\approx 85~\mu$K
at $l\approx 200$ and a fall at $l>300$. The data 
strongly disfavor models with a peak in the spectrum 
at $l=400$. Future work will include the analysis of the
\tocoB\ HEMT and remaining \D-band data.

\acknowledgments

We gratefully acknowledge conversations with and help from
Dave Wilkinson, Norm Jarosik, Suzanne Staggs, Steve Myers,
David Spergel, Max Tegmark, Angel Ot\'arola, Hern\'an Quintana,
the Princeton Machine Shop, Bernard Jones, Harvey Chapman,
Stuart Bradley, and Eugenio Ortiz.
The experiment would not have been possible
without NRAO's site monitoring and detector development.  We also
thank Lucent Technologies for donating the radar trailer.  This
work was supported by an NSF NYI award, a Cottrell Award from the
Research Corporation,
a David and Lucile Packard Fellowship (to L.~P.),
an NSF Career award (AST-9732960, to M.~D.),
a NASA GSRP fellowship to A.~M.,
an NSF graduate fellowship to M.~N.,
a Robert H. Dicke fellowship to E.~T.,
NSF grants PHY-9222952, PHY-9600015, and the University of Pennsylvania.
The data will be made public upon publication of this {\it Letter}.

%%%%%%%%%%%%%%%%%%%%%%%%%%%%%%%%%%%%%%%%%%%%%%%%%%%%%%%%%%%%%%%%%%%%%%%%%%%%%%%
% Bibliography

%%%%%%%%%%%%%%%%%%%%%%%%%%%%%%%%%%%%%%%%%%%%%%%%%%%%%%%%%%%%%%%%%%%%%%%%%%%%%%%
% Figure captions

%%%%%%%%%%%%%%%%%%%%%%%%%%%%%%%%%%%%%%%%%%%%%%%%%%%%%%%%%%%%%%%%%%%%%%%%%%%%%%%
% Table

\begin{deluxetable}{lcccccccc}
\footnotesize
\tablecaption{\tocoB\ Angular Spectrum \label{tbl:pspec}}
\tablewidth{18cm}
\tablehead{ &
\colhead{\Done} &
\colhead{\Done} &
\colhead{\Done} &
\colhead{\Dtwo} &
\colhead{\Dtwo} &
\colhead{\Dtwo} &
\colhead{\Done+\Dtwo} &
\colhead{\Done+\Dtwo} \nl
\colhead{$N_{\rm bins}$\tablenotemark{a}} &
\colhead{n-pt} &
\colhead{$l_{\rm eff}$\tablenotemark{b}} &
\colhead{$\delta T_l$\tablenotemark{c}} &
\colhead{n-pt} &
\colhead{$l_{\rm eff}$\tablenotemark{b}} &
\colhead{$\delta T_l$\tablenotemark{c}} &
\colhead{$l_{\rm eff}$\tablenotemark{b}} &
\colhead{$\delta T_l$\tablenotemark{c}} \nl
     &  & & $\mu$K & & & $\mu$K &  & $\mu$K }
\startdata
 128(84) &
	\dots & \dots & \dots &
	5 & $128^{+26}_{-33}$ & $55^{+18}_{-17}$ &
	$128^{+26}_{-33}$ & $55^{+18}_{-17}$ \nl
 128(84) &
	5,6 & $145^{+31}_{-39}$ & $93^{+14}_{-12}$ &
	6 & $162^{+24}_{-38}$ & $67^{+18}_{-17}$ &
	$152^{+26}_{-38}$ & $82^{+11}_{-11}$ \nl
 192(125) &
	7,8 & $218^{+23}_{-46}$ & $86^{+13}_{-13}$  &
	7-10 & $231^{+55}_{-63}$ & $86^{+9}_{-9}$ &
	$226^{+37}_{-56}$ & $83^{+7}_{-8}$  \nl
 256(165) &
	9-12 & $299^{+53}_{-74}$ & $89^{+11}_{-11}$ &
	11,12 & $329^{+21}_{-47}$ & $<80$ 95\% &
	$306^{+44}_{-59}$ & $70^{+10}_{-11}$ \nl
 384(250) &
	13-16 & $418^{+47}_{-77}$ & $<82$ 95\% &
	13-17 & $406^{+59}_{-60}$  & $<82$ 95\% &
	$409^{+42}_{-65}$ & $<67$ 95\%  \nl
\enddata
\tablenotetext{}{NOTE--- (a) The number of bins on the sky followed by, 
in parentheses, the number used in the analysis due to the
galactic/atmosphere cut. (b)
The range for $\ell_{\rm eff}$ denotes
the range for which the Knox filter exceeds $e^{-1/2}$ times
the peak value. (c) The
error on $\delta T_\ell=[\ell(\ell+1)C_\ell/2\pi]^{1/2}$ is
comprised of experimental uncertainty and sample variance and is $1\sigma$. The
calibration error is {\bf not} included. 
}
\end{deluxetable}

%%%%%%%%%%%%%%%%%%%%%%%%%%%%%%%%%%%%%%%%%%%%%%%%%%%%%%%%%%%%%%%%%%%%%%%%%%%%%%%
% The end

\end{document}